\documentclass[fleqn,12pt]{article}
 \usepackage{epsfig}
\usepackage{amssymb}
\def\cal{\fam2 }
\newcommand{\ba}{\begin{eqnarray}}
\newcommand{\re}{{\rm Re\,}}
\newcommand{\im}{{\rm Im\,}}
\newcommand{\ea}{\end{eqnarray}}
\newcommand{\be}{\begin{equation}}
\newcommand{\ee}{\end{equation}}

\newcommand{\eq}[1]{Eq.\,(\ref{#1})}

\newcommand{\greaterabout}{\raisebox{-.6ex}{\ $\stackrel{>}{\sim }$\ }}

\newcommand{\x}{(\nu/m)}
\newcommand{\y}{(\nu_0/m)}
\newcommand{\sigtot}{\sigma_{\rm tot}}

\newcommand{\betaP}{\beta_{\cal P'}}

\def\bea{\begin{eqnarray}} 
\def\eea{\end{eqnarray}} 
\def\rd{{\mathrm d}} 
 
\def\intd4x{\int{\rd}^4x}

\def\m32{{m_{3/2}}}

\def\sigevennu{\sigma_{\rm even}(\nu)}
\def\sigoddnu{\sigma_{\rm odd}(\nu)}

\begin{document}
\renewcommand\thepage{\ }
\begin{titlepage} 
%
\newcommand\reportnumber{1202} 
\newcommand\mydate{\today} 
\newlength{\nulogo} 
\settowidth{\nulogo}{\small\sf{NUHEP Report xxxx}}
\title{
\vspace{-.8in} 
\hfill\fbox{{\parbox{\nulogo}{\small\sf{
NUHEP Report  \reportnumber\\
          \mydate}}}}
\vspace{0.5in} \\
{
New analyticity constraints on the high energy behavior of hadron-hadron cross sections}}

\author{
M.~M.~Block\\
{\small\em Department of Physics and Astronomy,} \vspace{-5pt} \\ 
{\small\em Northwestern University, Evanston, IL 60208}\\
\vspace{-5pt}
\  \\
 \\
\vspace{-5pt}\\
%
\vspace{-5pt}\\
%
}    
\vspace{.5in}
\vfill
\date {}
\maketitle

\begin{abstract}
We here comment on a series of recent papers by Igi and Ishida[K. Igi and M. Ishida, Phys. Lett B {\bf 622}, 286 (2005)] and Block and Halzen[M. M. Block and F. Halzen, Phys. Rev D {\bf 72}, 036006 (2005)] that fit high energy $pp$ and $\bar pp$ cross section and $\rho$-value data, where $\rho$ is the ratio of the real to the imaginary portion of the forward scattering amplitude. These authors used Finite Energy Sum Rules and analyticity consistency conditions, respectively, to constrain the asymptotic behavior of hadron cross sections by anchoring their high energy asymptotic amplitudes---even under crossing---to low energy experimental data. Using analyticity, we here show  that i) the two apparently very different approaches are in fact equivalent, ii) that these analyticity constraints can be extended to give new constraints, and iii) that these constraints can be extended to crossing odd amplitudes. We also apply these extensions  to photoproduction. A new interpretation of duality is given.

\end{abstract}
\end{titlepage} 
\renewcommand{\thepage}{\arabic{page}} 
 About 40 years ago, Dolen, Horn and Schmid\cite{dolen-horn-schmid} used analyticity to derive finite-energy sum rules, FESRs, to determine Regge parameters (for  what were then high energies)  from low-energy data.
Very recently, Igi and Ishida, again using analyticity,  developed FESRs for both pion-proton scattering\cite{igiandishidapip}  and for $pp$ and $\bar pp$\ scattering\cite{igi} for rising cross sections at  present day energies.   They exploited the very  precise experimental cross section information, $\sigtot(pp)$ and $\sigtot(\bar pp)$, available for {\em low energy} scattering, to constrain the  coefficients of a real analytic amplitude fit they made to the even (under crossing) cross section $\sigma_+(\nu)$ at {\em high energies}. Block and Halzen{\cite{{bh},{newfroissart}}, taking  a very different approach, required that {\rm both} the hh (hadron-hadron) and the $\bar{\rm h}$h  low energy cross sections constrain the high energy fit, using
\ba
\sigtot({\nu_0})&=& \tilde \sigma({\nu_0})\qquad{\rm and}\qquad
\frac{d\sigtot}{d\nu  }({\nu_0})\quad=\quad\frac{d\tilde \sigma}{d\nu \ }({\nu_0}),\nonumber
\ea
where 
$\sigtot({\nu_0})$ is the {\em experimental}\,  hh or $\bar {\rm h}$h total cross section at laboratory energy ${\nu_0}$ and  $ \tilde \sigma({\nu_0})$ is the total cross section at $\nu_0$ obtained from the {\em high energy parametrization} that was  fit  the high energy  hh or $\bar {\rm h}$h cross section data for hadron-hadron scattering; both even and odd amplitudes (under crossing) were used. In the above, the transition energy $\nu_0$ was chosen to be an energy just above the resonance region, where the cross section energy dependence is smooth and featureless. In particular, they successfully fit $\gamma p$\cite{bh} and separately, $\pi^+p , \pi^-p$ and $pp,\ \bar pp$ scattering\cite{newfroissart} with a $\ln^2 s$ parametrization. In a separate work\cite{FESR}, they showed that they got identical numerical  results using these constraints as they got from using the Igi and Ishida constraint\cite{igi}, when 
fitting the {\em same} data set of $pp$ and $\bar pp$ high energy cross  sections. We will show below that the two approaches  are  equivalent, with both following from analyticity requirements.
 
In deriving their FESR2 for $pp$ and $\bar pp$ scattering\cite{igi}, Igi and Ishida\cite{igi} took a slightly different philosophy from Dolen, Horn and Schmid\cite{dolen-horn-schmid} in that they used terms for the high energy behavior that involved   non-Regge amplitudes such as terms in $\ln s$ and $\ln^2 s$, in addition to the Regge poles of ref. \cite{dolen-horn-schmid}.   They chose for their crossing-even high energy forward scattering amplitude%
\footnote{We have changed their notation slightly, replacing the amplitude $F$ by $f$, and the energy $N$ by $\nu_0$. In what follows, $m$ is the proton mass, $p$ is the laboratory momentum and $\nu$ is the laboratory energy. We have changed their notation for their dimensionless parameters, letting $c_0\rightarrow C_0$, $c_1\rightarrow C_1$, $c_2\rightarrow C_2$ and $\betaP\rightarrow B_{\cal P'}$.} %
$\tilde f_+(\nu)$, ($\tilde f_+(-\nu)=\tilde f_+(\nu)$), 
\ba
\im \tilde f_+(\nu)&=&\frac{\nu}{m^2}\left[C_0+C_1\ln\left(\frac{\nu}{m}\right)+C_2\ln^2\left(\frac{\nu}{m}\right)+B_{\cal P'}\left(\frac{\nu}{m}\right)^{\mu-1}\right],\label{imagf+}\\
\re \tilde f_+(\nu)&=&\frac{\nu}{m^2}\left[\frac{\pi}{2}C_1+C_2\pi \ln\left(\frac{\nu}{m}\right)-B_{\cal P'}\cot\left({\pi\mu\over 2}\right)\left(\frac{\nu}{m}\right)^{\mu -1}\right],\label{realf+}
\ea
where $m$ is the proton mass and $\nu$ is the laboratory projectile energy, with real dimensionless coefficients $C_0,\ C_1$, $C_2$ and $B_{\cal P'}$.

We comment that had they used the factor $p/m^2$ rather than $\nu/m^2$ in front of the right-hand sides  of \eq{imagf+} and \eq{realf+} as required by analyticity for a real {\em even} amplitude, their choice of amplitude would have been an even real analytic function and $f_+(\nu)$ would be zero for $0\le\nu\le m$, the proton mass, as required for a real analytic forward scattering amplitude (see ref. \cite{bc}). In the high energy limit---in \eq{imagf+} and \eq{realf+}---they replaced the laboratory momentum $p=\sqrt{\nu^2-m^2}$ by $\nu$. Using the optical theorem, after letting $p\rightarrow\nu$, they obtained the even cross section  from \eq{imagf+} as
\be
\tilde\sigma_+(\nu)=\frac{4\pi}{m^2} \left[C_0+C_1\ln(\nu/m)+C_2\ln^2(\nu/m)+B_{\cal P'}(\nu/m)^{\mu-1}\right], \label{sig+}
\ee
valid in the high energy region  $\nu \greaterabout {\nu_0}$. 
They used a  Reggion trajectory with $\mu=0.5$. 

Block and Cahn\cite{newfroissart} used a similar parametrization to analyze $pp$ and $\bar pp$ cross sections and $\rho$-values.  
Their  even real analytic forward high energy scattering  amplitude $\tilde f_+(\nu)$ is  given by: 
\ba
\im \tilde f_+(\nu)&=&\frac{p}{4\pi}\left[c_0+c_1\ln\left(\frac{\nu}{m}\right)+c_2\ln^2\left(\frac{\nu}{m}\right)+\betaP\left(\frac{\nu}{m}\right)^{\mu-1}\right]\quad{\rm for }\ \nu\ge m,\nonumber\\
\im \tilde f_+(\nu)&=&0\qquad\qquad\qquad\qquad\qquad\qquad\qquad\qquad\qquad\qquad\qquad{\rm for }\ 0\le\nu\le m,\label{imagf+bh}\\
\re \tilde f_+(\nu)&=&\frac{p}{4\pi}\left[\frac{\pi}{2}c_1+c_2\pi \ln\left(\frac{\nu}{m}\right)-\beta_{\cal P'}\cot\left({\pi\mu\over 2}\right)\left(\frac{\nu}{m}\right)^{\mu -1}\right].\label{realf+bh}
\ea 
Using the optical theorem, their even cross section is 
\be
\tilde\sigma_+(\nu)=c_0+c_1\ln(\nu/m)+c_2\ln^2(\nu/m)+\betaP(\nu/m)^{\mu-1},\label{sig+bh}
\ee
where here the coefficients $c_0$, $c_1,\ c_2$ and $\betaP$ have dimensions of mb.  

We now introduce $f_+(\nu)$, the {\em true} even forward scattering amplitude (which of course, we do not know!), valid for all $\nu$, where $f_+(\nu)\equiv[f_{pp}(\nu)+f_{\bar pp}(\nu)]/2$, using forward scattering amplitudes for $pp$ and $\bar pp$ collisions. 
Using the optical theorem, the imaginary portion of $f_+(\nu)$  is related to the  even  total cross section $\sigevennu$ by
\ba
\im f_+(\nu)&=&\frac{p}{4\pi}\sigevennu \!\qquad\mbox{for $\nu\ge m$}.
\label{sig+1}
\ea

Next, define the  odd amplitude $\nu\hat f_+(\nu)$ as the difference 
\be
\nu\hat f_+(\nu)\equiv\nu\left[f_+(\nu)-\tilde f_+(\nu)\right]\label{fsuper}
,
\ee
which satisfies the unsubtracted odd amplitude  dispersion relation
\be
\re \nu \hat f_+(\nu)=\frac{2\nu}{\pi}\int^\infty_0\frac{\im \nu\,' \hat f_+(\nu\,')}{\nu\,'^2-\nu^2}\,d\nu\,'. \label{dispersion}
\ee
Since for large $\nu$, the odd amplitude $\nu \hat f_+(\nu)\sim\nu^\alpha$ ($\alpha<0$) by design, it also satisfies the super-convergence relation
\be
\int_0^\infty{\rm Im}\,\nu\hat f_+(\nu)\,d\nu=0.\label{superconvergence}
\ee

In ref. \cite{dolen-horn-schmid}, the FESRs are given by
\be
\int_0^{\nu_0}\nu^n\,{\rm Im}\hat f \,d\nu=\sum \frac {{\nu_0}^{\alpha +n+1}}{\alpha + n+1},\qquad n= 0,1,\ldots,\infty, \label{FESRHorn}
\ee
where $\hat f(\nu)$ is crossing-even for odd integer $n$ and crossing-odd for even integer $n$. 
In analogy to the $n=1$ FESR of ref. \cite{dolen-horn-schmid}, which requires the odd amplitude $\nu \hat f(\nu)$, Igi and Ishida inserted the super-convergent  amplitude of \eq{fsuper} into the super-convergent dispersion relation of \eq{superconvergence}, obtaining 
\be
\int_0^{\infty}\nu \,\im \left[f_+(\nu)- \tilde f_+(\nu)\right]\, d\nu.\label{super0}
\ee
We note that the odd difference amplitude $\nu \im  \hat f_+(\nu)$ satisfies \eq{superconvergence}, a super-convergent dispersion relation, even if neither $\nu\, \im f_+(\nu)$ nor $\nu \,\im \tilde f_+(\nu)$ satisfies it.  Since the integrand  of \eq{super0}, $\nu \,\im \left[f_+(\nu)- \tilde f_+(\nu)\right]$,  is super-convergent, we can truncate the upper limit of the integration at the finite energy $\nu_0$, an energy high enough for resonance behavior to vanish and where the difference between the two amplitudes---the true amplitude $f_+(\nu)$ minus $\tilde f_+(\nu)$, the amplitude which parametrizes the high energy behavior---becomes negligible, so that the integrand can be neglected for energies greater than $\nu_0$. Thus, after some rearrangement, we get the even finite energy sum rule (FESR)
\be
\int_0^{\nu_0}\nu \im f_+(\nu)\,d\nu=\int_0^{\nu_0}\nu\im \tilde f_+(\nu)\, d\nu.
\label{truef}
\ee 

Next, the left-hand integral of \eq{truef} is broken up into two parts, an integral from $0$ to $m$ (the `unphysical' region) and the integral from $m$ to ${\nu_0}$, the physical  region. We use the optical theorem to evaluate the left-hand integrand for $\nu\ge m$. After noting that the imaginary portion of $\tilde f_+(\nu)=0$ for $0\le\nu\le m$, we again use the optical theorem  to evaluate the right-hand integrand, finally obtaining  the finite energy sum rule FESR(2) of Igi and Ishida, in the form: 
\ba
\int_0^m\nu\, \im f_+(\nu)\,d\nu+\frac{1}{4\pi}\int_m^{{\nu_0}}\nu p\,\sigma_{\rm even}(\nu)\,d\nu&=&\frac{1}{4\pi}\int_m^{\nu_0}\nu p\,\tilde\sigma_+(\nu)\, d\nu.\label{intftilde}
\ea

We now enlarge on the consequences of \eq{intftilde}. We note that if \eq{intftilde} is valid at the upper limit ${\nu_0}$, it certainly is also valid at ${\nu_0}+\Delta {\nu_0}$, where  $\Delta {\nu_0}$ is very small compared to ${\nu_0}$, i.e., $0\le \Delta \nu_0\ll\nu_0$. Evaluating \eq{intftilde2} at the energy ${\nu_0}+\Delta {\nu_0}$ and then subtracting \eq{intftilde} evaluated at ${\nu_0}$, we find
\be
\frac{1}{4\pi}\int_{\nu_0}^{{\nu_0}+\Delta {\nu_0}} \nu p\,\sigevennu\,d\nu=\frac{1}{4\pi}\int_{\nu_0}^{{\nu_0}+\Delta {\nu_0} }\nu p\tilde\sigma_+(\nu)\,d\nu.\label{FESR(2)_4}
\ee

Clearly, in the limit of $\Delta {\nu_0}\rightarrow 0$, \eq{FESR(2)_4} goes into
\be
\sigma_{\rm even}({\nu_0})=\tilde\sigma_+({\nu_0}).\label{sig+=sighigh}
\ee
Obviously, \eq{sig+=sighigh} also implies that
\be
\sigma_{\rm even}({\nu})=\tilde\sigma_+({\nu})\qquad\mbox{\rm for all $\nu\ge \nu_0$},\label{sig+=sighigh1}
\ee
but is most useful in practice when $\nu_0$ is as low as possible. The utility of \eq{sig+=sighigh1} becomes evident when we recognize that the left-hand side of it can be evaluated using  the very accurate low energy {\em experimental} crossing-even total cross section data, whereas the right-hand side can use the phenomenologist's parametrization of the {\em high} energy cross section. For example, we could use the cross section parametrization of \eq{sig+bh} on the right-hand side of \eq{sig+=sighigh1} and write the constraint
\be
\left[\sigma_{pp}(\nu)+\sigma_{\bar pp}(\nu)\right]/2=c_0+c_1\ln(\nu/m)+c_2\ln^2(\nu/m)+\betaP(\nu/m)^{\mu-1},\label{sig+bh1}
\ee
where $\sigma_{pp}$ and $\sigma_{\bar pp}(\nu)$ are the {\em experimental} $pp$ and $\bar pp$ cross sections at the laboratory energy $\nu$.
Equation (\ref{sig+=sighigh}) (or \eq{sig+=sighigh1}) is our first important extension, giving us  an analyticity constraint, a consistency condition  that the even high energy (asymptotic) amplitude must satisfy.  

Reiterating, \eq{sig+=sighigh1} is a consistency condition imposed by analyticity that states that we must fix the even high energy cross section  evaluated at energy $\nu\ge {\nu_0}$ (using the asymptotic even amplitude) to the low energy {\rm experimental} even cross section at the {\em same}  energy $\nu$, where $\nu_0$ is an energy just above the resonances. Clearly, \eq{sig+=sighigh} also implies that all derivatives of the total cross sections match, as well as the cross sections themselves, i.e., 
\ba
\frac{d^n\sigma_{\rm even}}{d\nu^n\ \ \ \  }({\nu})&=& \frac{d^n\tilde\sigma_+}{d\nu^n\ }({\nu}),\quad n=0,1,2,\ldots\qquad\nu\ge \nu_0\label{derivative}
\ea
giving new even amplitude  analyticity constraints. Of course, the evaluation of \eq{derivative} for $n=0$ and $n=1$ is effectively the same as evaluating \eq{derivative} for $n=0$ at two nearby values, $\nu_0$ and $\nu_1>\nu_0$. It is up to the phenomenologist to decide which {\em experimental set}\, of quantities it is easier to evaluate.

We emphasize that these consistency constraints  are the consequences of imposing analyticity, implying several important conditions:
\begin{enumerate}
\item 
The new constraints that are derived here tie together both the even  hh and $\bar {\rm h}$h experimental cross sections and their derivatives to the even high energy approximation that is used to fit data at energies well above the resonance region. Analyticity then requires that there  should be a {\em good} fit to the high energy data {\em after} using these constraints, i.e., the $\chi^2$ per degree of freedom of the constrained fit should be $\sim 1$, {\em if}\, the high energy asymptotic amplitude is a  good approximation to the high energy data. This is our consistency condition demanded by analyticity.  If, on the other hand,  the high energy asymptotic amplitude would have given a somewhat poorer fit to the data when {\em not} using the new constraints, the effect is tremendously magnified by utilizing  these new constraints, yielding a very large $\chi^2$ per degree of freedom.  As an example, both Block and Halzen\cite{bh} and Igi and Ishida\cite{igiandishidapip,igi} {\em conclusively} rule out  a $\ln s$ fit to both $\pi^\pm p$ and $pp$ and $\bar pp$ cross sections and $\rho$-values because it has a huge $\chi^2$ per degree of freedom.\label{point1}
\item  Consistency with analyticity  requires that the results be valid for all $\nu\ge {\nu_0}$, so that the constraint doesn't depend on the particular choice of $\nu$.\label{point2} 
\item No evaluation of  the non-physical integral $\int_0^m\nu\, \im f_+(\nu)\,d\nu$ used in \eq{intftilde} is needed for our new constraints. Thus, the exact value of non-physical integrals, even if very large, does not affect our new constraints.\label{point3}
\item As stated before, $\nu_0$ is an energy slightly above the resonance region where the energy behavior of the cross section is smooth and featureless.  Duality previously has been used to state that the average value of the energy moments of the imaginary portion of the  true amplitude over the energy interval 
0 to ${\nu_0}$ are the same as the average value of the energy moments of the high energy approximation amplitude over the same interval\cite{dolen-horn-schmid}, which is illustrated in Fig. \ref{fig:resonances}. 

Here, we present a new interpretation of duality---we have  demonstrated  that analyticity requires that the even cross sections and their derivatives deduced from the even dual high energy amplitude at energy $\nu_0$ are the same as those cross section and their derivatives found from low energy experimental cross section data at $\nu_0$, under the caveat that  the dual amplitude gives a good representation of the high energy data. Later, we will demonstrate that this is also true for odd amplitudes, so that our new duality  interpretation is  true for hh and $\bar {\rm h}$h cross sections, as well. \label{point4}  
\end{enumerate} 

Having restricted ourselves so far to 
even  amplitudes, let us now consider odd amplitudes, defining $f_-(\nu)$ as the {\em true}\,  odd forward scattering amplitude, valid for all $\nu$ (again, which we do not know!). In terms of the forward scattering amplitudes for $pp$ and $\bar pp$ collisions, we define
$f_-(\nu)\equiv[f_{pp}(\nu)-f_{\bar pp}(\nu)]/2.$ Using the optical theorem, the imaginary portion of the odd amplitude is related to the physical odd (under crossing)  total cross section $\sigma_{\rm odd}$ by
\be
\im f_-(\nu)=\frac{p}{4\pi}\sigoddnu,\qquad\mbox{for $\nu\ge m$.}\label{sig-1}
\ee 

We now define a new super-convergent odd amplitude $\hat f_-(\nu)$ as
\ba
\hat f_-(\nu)&\equiv& f_-(\nu)-\tilde f_-(\nu)
,\label{fhat2}
\ea
where $\tilde f_-(\nu)$ is our high energy parametrization amplitude, related to the odd (under crossing) high energy cross section $\sigma_-(\nu)$ by
\ba
\im f_-(\nu)&=&\frac{p}{4\pi}\sigma_-(\nu),\quad\mbox{for }\nu\ge m\nonumber\\
\im f_-(\nu)&=&0\qquad\qquad\quad 0\le \nu \le m.\label{fodd2}
\ea

The super-convergent amplitude of \eq{fhat2} satisfies the unsubtracted odd amplitude dispersion relation
\be
\hat f_-(\nu)=\frac{2\nu}{\pi}\int_0^\infty\frac{{\rm Im}\,\hat f_-(\nu')}{\nu'^2-\nu^2}\,d\nu',
\ee
and, as before, it also satisfies the super-convergent dispersion relation
\be
\int_0^\infty{\rm Im}\,\hat f_-(\nu)\,d\nu=0.\label{superconvergence-}
\ee
Again, we can truncate the integral at $\nu_0$, so that
\be
\int_0^{\nu_0}{\rm Im}\,\hat f_-(\nu)\,d\nu=0,\label{superconvergence-2}
\ee
or 
\be
\int_0^{\nu_0}{\rm Im}\,f_-(\nu)\,d\nu=\int_0^{\nu_0}{\rm Im}\,\tilde f_-(\nu)\,d\nu.\label{superconvergence-3}
\ee
After applying  the optical theorem, using \eq{sig-1} on the left-hand side and  \eq{fodd2}  on the right-hand side of \eq{superconvergence-3},  we write our new $n=0$  odd finite energy sum rule called FESR(odd) as
\ba
\int_0^m\im f_-(\nu)\,d\nu+\frac{1}{4\pi}\int_m^{{\nu_0}} p\sigma_{\rm odd}(\nu)\,d\nu&=&\frac{1}{4\pi}\int_m^{\nu_0}p\tilde \sigma_-\, d\nu,\qquad {\rm FESR(odd).}\label{intftilde2}
\ea

Following the same line as before, it is straightforward to show for  odd amplitudes that FESR(odd) implies that
\ba
\frac{d^n\sigma_{\rm odd}}{d\nu^n \ \  }({\nu})&=& \frac{d^n\tilde\sigma_-}{d\nu^n }({\nu}),\quad n=0,1,2,\ldots,\qquad \nu \ge \nu_0,\label{oddconstraints}
\ea
where $\tilde\sigma_-(\nu)$ is the odd (under crossing) high energy cross section approximation and $\sigma_{\rm odd}(\nu)$ is the experimental odd cross section.

Thus, we have now derived new analyticity constraints for {\em both}
even and odd cross sections, allowing us to constrain both $\rm{hh}$ and
$\bar{\rm h}$h scattering. All of the Conditions \ref{point1}, \ref{point2}, \ref{point3} and  \ref{point4}, enumerated earlier for even amplitudes, are now valid for odd amplitudes, and hence, for both $\rm{hh}$ and $\bar{\rm h}$h scattering.

Block and Halzen\cite{newfroissart} expanded upon these ideas, using linear combinations of cross sections and derivatives to anchor {\em both} even  and odd cross sections.  A total of 4 constraints, 2 even and 2 odd constraints, were used by them in their successful $\ln^2s$ fit to $pp$ and $\bar pp$\  cross sections and $\rho$-values, where they first did a local fit to $pp$ and $\bar pp$\ cross sections and their slopes in the neighborhood of $\nu_0=7.59$ GeV (corresponding to $\sqrt s_0=4$ GeV), to determine the experimental cross sections and their first derivatives at which  they anchored their fit. The data they used in the high energy fit were  $pp$ and $\bar pp$ cross sections and $\rho$-values with energies $\sqrt s\ge 6$ GeV. Introducing the even  cross section $\sigma_0(\nu)$, they parametrized the high energy cross sections and $\rho$ values\cite{newfroissart} as 
\begin{eqnarray}
\sigma_0(\nu)&{\!\!\! =\!\!\! }&c_0+c_1\ln\left(\frac{\nu}{m}\right)+c_2\ln^2\left(\frac{\nu}{m}\right)+\beta_{\cal P'}\left(\frac{\nu}{m}\right)^{\mu -1}\label{sigma0},\\
\sigma^\pm(\nu)&{\!\!\! =\!\!\! }&\sigma_0\left(\frac{\nu}{m}\right)
\pm\  \delta\left({\nu\over m}\right)^{\alpha -1},\label{sigmapm}\\
\rho^\pm(\nu)&{\!\!\! =\!\!\! }&{1\over\sigma^\pm}\left\{\frac{\pi}{2}c_1+c_2\pi \ln\left(\frac{\nu}{m}\right)-\beta_{\cal P'}\cot\left({\pi\mu\over 2}\right)\left(\frac{\nu}{m}\right)^{\mu -1}+\frac{4\pi}{\nu}f_+(0)\right.\nonumber\\
&&\qquad\qquad\left.\pm\  \delta\tan\left({\pi\alpha\over 2}\right)\left({\nu\over m}\right)^{\alpha -1} \right\}\!\!.\label{rhopm}
\end{eqnarray}
We note that the even coefficients $c_0, c_1, c_2$ and $\beta_{\cal P'}$ are the same as those used in \eq{sig+bh1}. The real constant $f_+(0)$ is the subtraction constant\cite{{bc},{gilman}} required at $\nu=0$ for a singly-subtracted dispersion relation.  They also  used $\mu=0.5$. 
The odd cross section in \eq{sigmapm} is given by
\be
\delta\left({\nu\over m}\right)^{\alpha -1},\label{odd1}
\ee
described by two parameters, the coefficient $\delta$ and the Regge power $\alpha$. 
We define 
\begin{eqnarray}
\Delta\sigma(\nu_0)&\equiv&\frac{\sigma^{+}\y-\sigma^-\y}{2}\qquad\quad\quad
=\,\quad\delta\y^{\alpha -1},\label{derivodd}\\
\Delta m(\nu_0)&\equiv&\frac{1}{2}\left(\frac{d\sigma^{+}}{d\x}-\frac{d\sigma^{-}}{d\x}\right)_{\nu =\nu_0}\quad\quad
=\quad\delta\left\{(\alpha -1)\y^{\alpha - 2}\right\},\label{interceptodd}
\end{eqnarray}
in terms of the  odd experimental values at  $\nu_0$.  Since now $\delta$ and $\alpha$ are completely fixed by the experimental quantities $\Delta \sigma(\nu_0)$ and $\Delta m(\nu_0)$, these two analytic constraints severely restricts the phenomenologist using this particular choice of amplitude.  If  \eq{odd1} is not a particularly good  representation of the high energy data, the $\chi^2$ from the  fit will be very poor.  On the other hand, if the $\chi^2$ is very good---as found by Block and Halzen\cite{newfroissart}---it provides great confidence in the choice of \eq{odd1} as the imaginary portion of the asymptotic odd amplitude.

Finally, to get a physical picture of what the new analyticity constraints look like compared to FESR(2), we apply  \eq{sigma0}, the even cross section,  to spin-averaged $\gamma p$ scattering. For the $\gamma p$ system, the left-hand integral of \eq{intftilde}, involving  experimental cross sections $\sigma_{\gamma p}(\nu)$ in the resonance regions ($\nu\le \nu_0$), is now 
$\int_0^{\nu_0} \nu^2 \sigma_{\gamma p}(\nu)\,d\nu.$ As a function of the center-of-mass energy $\sqrt s$, Fig. \ref{fig:resonances} shows 3 separate plots: the experimental resonance cross section data multiplied by $\nu^2$ as the open circles; the $\sigma_{\gamma p}({\rm fit})$---a fit to the resonance region made by Damashek and Gilman \cite{gilman2}---after multiplication by $\nu^2$, as the dashed-dot curve; and, finally, the cross section $\sigma_0\times\nu^2$,  where $\sigma_0$ is the even high energy cross section parametrization from  Block and Halzen\cite{bh}, as the dashed curve. Block and Halzen used a transition energy $\nu_0=1.68$ GeV ($\sqrt s_0=2.01$ GeV) in their fit,  requiring that their fit  match the experimental cross section and first derivative at $\nu_0$. Note that the resonance data oscillate about the smooth high energy fit, with the oscillations gradually damping out so that experimental data approach the high energy fit as we near the transition energy of $\sqrt s_0=2.01$ GeV.

In conclusion, we now have new analyticity constraints for both even and odd amplitudes---additional tools for the phenomenologist to use in fitting hadron-hadron total cross sections and $\rho$-values. In practice, these new analyticity constraints are much simpler to use than FESRs. The fits are  anchored  by the experimental cross section data near the transition energy $\nu_0$, so neither complicated numerical integrations nor evaluation of  unphysical regions are required. These consistency constraints are due to the application of analyticity to finite energy integrals---the analog of analyticity giving rise to traditional dispersion relations when it is applied to integrals with infinite upper limits---giving us a new interpretation of the duality principle.

\begin{figure}[h] 
\begin{center}
\mbox{\epsfig{file=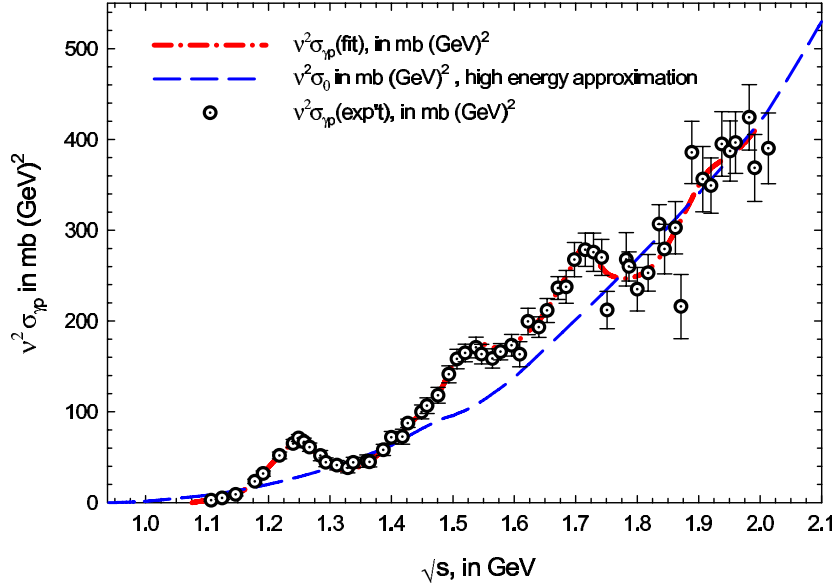,width=4.5in%
,bbllx=65pt,bblly=240pt,bburx=500pt,bbury=545pt,clip=%
}}
\end{center}
\caption[]{ \footnotesize 
 The integrands of the FESR2 rule. The open circles are $\nu^2\times \sigma_{\gamma p}{\rm (exp't)}$, the dash-dotted curve is $\nu^2\times \sigma_{\gamma p}({\rm fit})$, and the dashed curve is $\nu^2\times\sigma_0$, all in mb (GeV)$^2$ {\em vs.} $\sqrt s$, in GeV.   $\sigma_0=c_0+c_1\ln(\nu/m)+\ln^2(\nu/m)+\beta_{\cal P'}(\nu/m)^{-.5}$ is the theoretical high energy fit  of Block and Halzen\cite{bh}, $\sigma_{\gamma p}({\rm fit})$ is the  resonance cross section fit  of Damashek and Gilman\cite{gilman2}, and $ \sigma_{\gamma p}{\rm (exp't)}$ are the experimental data in the resonance region. The transition energy  was $\sqrt s_0=2.01$ GeV.

  }
\label{fig:resonances}
\end{figure}



\begin{thebibliography}{00}
\bibitem{gilman} 
For the reaction $\gamma +p\rightarrow\gamma + p$, it is fixed as the Thompson scattering limit $f_+(0)=-\alpha/m=-3.03\  \mu {\rm b\  GeV}$:  see ref. \cite{gilman2}.
\end{thebibliography}
\end{document}